\documentstyle[aps,prd,twocolumn,eqsecnum,epsfig,floats,float]{revtex}

\input{amssym.def}
\newsymbol\lesssim 132E
\newsymbol\gtrsim 1326

\begin{document}

\title{Critical phenomena and a new class of self-similar spherically
  symmetric perfect-fluid solutions}

\author{
  B. J. Carr,${}^{1}$\thanks{E-mail: B.J.Carr@qmw.ac.uk}
  A. A. Coley,${}^{2}$\thanks{E-mail: aac@mscs.dal.ca}
  M. Goliath,${}^{3}$\thanks{E-mail: goliath@physto.se}
  U. S. Nilsson,${}^{4}$\thanks{E-mail:
unilsson@mercator.math.uwaterloo.ca}
  and C. Uggla${}^{5}$\thanks{E-mail: uggla@physto.se}}

\address{${}^{1}$Astronomy Unit, Queen Mary and Westfield College,
  University of London, Mile End Road, London E1 4NS, England}
\address{${}^{2}$Department of Mathematics and Statistics, Dalhousie
  University, Halifax, Nova Scotia, B3H 3J5, Canada} 
\address{${}^{3}$Department of Physics, Stockholm University,
  Box 6730, S-113 85 Stockholm, Sweden}
\address{${}^{4}$Department of Applied Mathematics, University of
Waterloo,
  Waterloo, Ontario N2L 3G1, Canada}
\address{${}^{5}$Department of Engineering Sciences, Physics and
  Mathematics, University of Karlstad, S-651 88 Karlstad, Sweden}

\date{\today}

\maketitle

\begin{abstract}
  We consider the self-similar solutions associated with the critical 
  behavior observed in the gravitational collapse of spherically symmetric 
  perfect fluids with equation of state $p=\alpha\mu$. We identify for
  the first time the global nature of these solutions and show that it
  is sensitive to the value of $\alpha$. In particular, for
  $\alpha>0.28$, we show that the critical solution is associated with
  a new class of asymptotically Minkowski self-similar spacetimes. We
  discuss some of the implications of this for critical phenomena.
\end{abstract}

\pacs{0420, 0420J, 0440N, 9530S, 9880H}

\section{Introduction}

One of the most exciting developments in general relativity in recent
years has been the discovery of critical phenomena in gravitational
collapse. For a variety of spherically symmetric imploding matter
fields, there is a self-similar critical solution containing a naked
singularity which separates models which collapse to black holes from
those which disperse 
\cite{Choptuik1993,AbrahamsEvans1993,Choptuik1994}. Sometimes
a discrete similarity is involved but, in other circumstances, the
critical solution seems to be represented by a continuously
self-similar model. This is one which has a homothetic Killing vector
and contains no dimensional constants. Perfect-fluid models of this
kind necessarily have an equation of state of the form $p=\alpha\mu$
and so only in this case could the critical solution be homothetic.

Self-similar spherically symmetric perfect-fluid solutions have been much 
studied in general relativity (see \cite{CarrColey1998SS} and
references therein) and the attempt to understand critical phenomena
has led to a several further studies 
\cite{EvansColeman1994,Gundlach1995,Maison1996,Koike-et-al1999,Frolov1997}.
However, their precise relationship with the critical solution has
remained obscure. This is mainly because the full family of such
solutions had not been identified when critical phenomena were first
discovered. However, recently Carr \& Coley \cite{CarrColey1999classi}
have presented a complete asymptotic classification of such
solutions. Furthermore, by reformulating the field equations for these
models in terms of dynamical systems, Goliath et al.
\cite{Goliath-et-al1998SSS,Goliath-et-al1998TSS} have obtained a
compact three-dimensional state space representation of the solutions
and this leads to another complete picture of the solution
space. These investigations have resulted in the discovery of a new
class of `asymptotically Minkowski' self-similar spacetimes. 

In this paper we shall discuss these new solutions and show why they are
intimately related to critical phenomena \cite{EvansColeman1994}. We
thereby demonstrate for the first time the {\it global} nature of the
critical solution. Although the detailed derivation of these new
solutions is given elsewhere, this is the first published announcement
of their existence and the first attempt to link them to critical
phenomena. The purpose of this paper is therefore to highlight this
result in advance of the more extensive analyses (since these are in a
much broader context). In particular, we will show how the global
features relates to the equation of state parameter $\alpha$ and
explain why there is only one such solution for each $\alpha$. It
should be emphasized that numerical studies of the critical solution
are always restricted to some finite range of the self-similar
variable $z$. However, as the critical index is approached, the extent 
of the self-similar region grows and one could in principle go to
arbitrarily large values of $z$. This paper can therefore be regarded as
predicting the characteristics of these solutions. 

The discussion will mainly be in terms of the compact state space
\cite{Goliath-et-al1998SSS,Goliath-et-al1998TSS} but, to extract
important physical features, some of the quantities used by
Carr \& Coley  \cite{CarrColey1999classi} will be plotted. No
equations will be used because the discussion is intended to be purely
qualitative and thereby accessible to the general reader. However,
some technical terms will be used in the next section, so we here give
some background references. For an introduction to dynamical systems
theory in general relativity, see \cite{book:WainwrightEllis1997}; for
its particular application in the spherically symmetric context, see
\cite{Goliath-et-al1998SSS,Goliath-et-al1998TSS}; for more details of
the other types of self-similar solutions, see
Carr et al. \cite{Carr-et-al1999}. At the end of the paper, we
emphasize our key predictions, so that ``critical'' workers can
investigate these. 

\section{The solution space}\label{sec:solu}

\begin{figure}
  \centerline{\hbox{\epsfig{figure=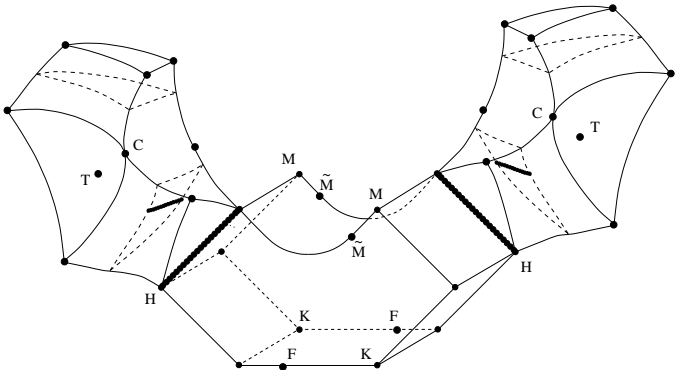,width=0.5\textwidth}}}
  \caption{The three-dimensional state space of self-similar spherically 
    symmetric perfect-fluid models for $\alpha>1/5$, obtained by
    matching a spatially self-similar region with two timelike
    self-similar regions along the lines H {\protect\cite{Carr-et-al1999}}.
    Labelled equilibrium points act as asymptotic states for
    orbits. The triangular surfaces indicated by dashed lines are
    sonic surfaces. Two of these surfaces contain a sonic line,
    indicated by a line of equilibrium points.}\label{fig:match} 
\end{figure}

We shall focus on spherically symmetric self-similar solutions in which
the spacetime admits a homothetic Killing vector. This means that all
dimensionless variables depend only on the self-similar variable
$z\equiv r/t$, where $r$ is the comoving radial coordinate and $t$ is
the time coordinate. In a dynamical systems approach these solutions
correspond to orbits in a three-dimensional compact state space. The
state space of self-similar spherically symmetric perfect-fluid models
(for $\alpha>1/5$) is presented in  Fig. \ref{fig:match}. A point in
this space corresponds to a certain geometry and matter field
configuration on a homothetic (constant $z$) slice, while an 
orbit in the state space represents an entire spacetime. All
continuous orbits are future and past asymptotic to one of a few
solutions with higher symmetry. These appear as equilibrium points on
the boundary of the state space and these points are labelled in
Fig. \ref{fig:match}. A physical description of the solutions
asymptotic to them is given in Table \ref{tab}. 

The state space is divided into two halves, one corresponding to positive 
$z$, the other to negative $z$. This means that the solutions in one half
are the time-reverse of solutions in the other half, so all
equilibrium points appear twice in Fig. \ref{fig:match}. The
{\em sonic surfaces} are also depicted and solutions generally develop a
shock-wave here \cite{CahillTaub}. However, in two of the sonic
surfaces there is a {\em sonic line} and solutions which pass through
this line can be extended continuously through the sonic surface. Only
these solutions will be considered to be physical and the number of
such solutions is strongly restricted. 

The state space has the advantage that it gives a pictorial
representation of the relationship between different solutions 
and the connection between the initial and final states, thereby
yielding insights into the global nature of the solutions. However, it
has the disadvantage that it is rather abstract. To better understand
the physical aspects of the solutions, it is useful to consider some
of the physically interesting quantities which arise in the comoving
approach. The dependence of these quantities on $z$ corresponds to
two-dimensional projections of orbits in the full state
space. Following \cite{CarrColey1999classi}, we use: 
(1) the scale factor $S$, which fixes the relation between the
comoving radial coordinate $r$ and the Schwarzschild radial coordinate
$R=Sr$, this indicating when a solution expands infinitely
($S\rightarrow \infty$) or encounters a singularity ($S\rightarrow 0$)
for finite values of $z$; 
(2) the velocity $V$ of the spheres of constant $z$ relative to the
fluid, which is important for the identification of event horizons
($|V|=1$) and naked singularities, see \cite{OriPiran1990};
(3) the density profile $\mu t^2$, which gives the matter distribution
at a given comoving time $t$; and
(4) the mass function  $2m/R$, where $m(r)$ is the mass within radial
coordinate $r$, this indicating the presence of an apparent horizon
($2m/R=1$), see \cite{HernandezMisner1966}. 

We shall first briefly review the previously known families of solutions.
All of these solutions are discussed in more detail in
\cite{Carr-et-al1999}, where the dependence of the above functions on
$z$ are shown explicitly. We shall then consider the new
asymptotically Minkowski solutions. This is the first discussion of
their physical features and the first detailed analysis of their
relevance to critical phenomena. 
\vspace*{3mm}

\noindent{\em Asymptotically Friedmann solutions}\\[0.5mm]
There are two one-parameter sets of solutions that are asymptotic to the
flat Friedmann solution, all of which are connected with one of the
Friedmann points F. One has positive $z$ and the other has negative
$z$. Two qualitatively different families can be distinguished: 
(1) expanding-recollapsing solutions (F--K orbits); and 
(2) ever-expanding (or ever-contracting) solutions (F--C orbits),
where C is to be interpreted as an infinitely dispersed state. The
latter family contains the flat Friedmann solution itself. Thus the
flat Friedmann solution appears both as an equilibrium point F and as
an orbit in state space, corresponding to different slicings. 

\begin{table}
  \begin{tabular}{@{}c@{\hspace{5mm}}l}
    Label & Interpretation \\ \hline
    C & Solutions with a regular center or \\
    & infinitely dispersed solutions \\
    M, $\tilde{\rm M}$ & Asymptotically Minkowski solutions \\
    K & Non-isotropic singularity solutions \\
    F & Asymptotically Friedmann solutions\\
    T & Exact static solution\\
  \end{tabular}
  \caption{Interpretation of solutions asymptotic to the given
    equilibrium points.}\label{tab} 
\end{table}

\vspace*{3mm}

\noindent{\em Asymptotically quasi-static solutions}\\[0.5mm]
For each value of $\alpha$, there is a unique static solution, originally 
found by Tolman \cite{Tolman1934}. The corresponding (T--T) orbit 
traverses the entire state space and spans both positive and negative $z$. 
Furthermore, there is a two-parameter set of solutions with behavior
resembling the static solution at large $|z|$ (i.e. at early
times). They are all associated with K points, corresponding to
non-isotropic singularities. As with the asymptotically Friedmann
solutions, there are two different families within this class: 
(1) expanding-recollapsing solutions (K--K orbits); and 
(2) ever-expanding (or ever-contracting) solutions (K--C orbits). The
latter contain the naked-singularity solutions discussed by
Ori \& Piran \cite{OriPiran1990} and Foglizzo \& Henriksen
\cite{FoglizzoHenriksen1993}. Unlike the asymptotically Friedmann 
solutions, the asymptotically quasi-static solutions necessarily span
both positive and negative $z$. 
\vspace*{3mm}

\begin{figure}
  \centerline{\hbox{\epsfig{figure=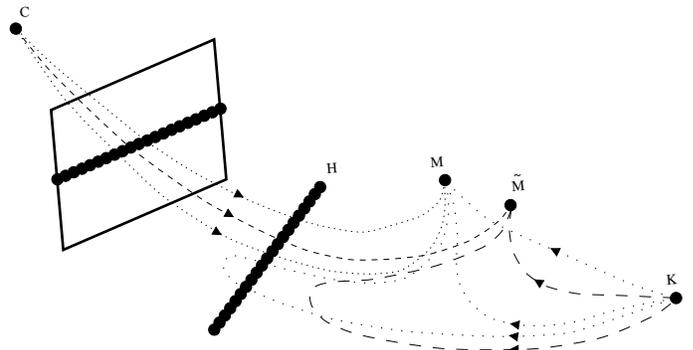,width=0.5\textwidth}}}
  \caption{Asymptotically Minkowski solutions of class A (dotted)
    and class B (dashed) as orbits in state space. Arrows go from C
    or K towards the infinitely dispersed state. Densely dotted and
    short-dashed curves correspond to regular solutions, while
    sparsely dotted and long-dashed curves correspond to singular
    solutions.}\label{fig:mgnu}  
\end{figure}

\begin{figure}
  \centerline{\hbox{\epsfig{figure=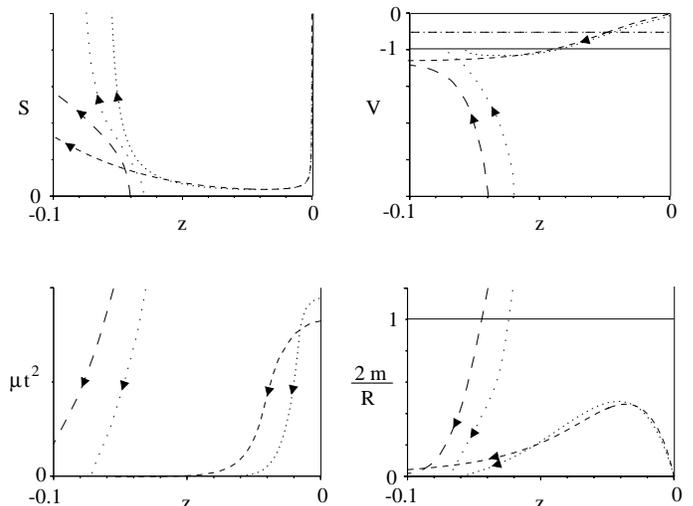,width=0.5\textwidth}}}
  \caption{Physical quantities for the asymptotically 
    Min\-kowski solutions. The dash-dotted line in the $V(z)$ diagram 
    corresponds to the sonic surface. Other designations are given in the
    caption of Fig. \ref{fig:mgnu}.}\label{fig:mcc}  
\end{figure}

\noindent{\em Asymptotically Minkowski solutions}\\[0.5mm]
When $\alpha>1/5$, solutions exist that are `asymptotically Minkowski', in 
the sense that the state-space orbits asymptote to equilibrium points that 
correspond to Minkowski space. There are actually two subclasses of such 
solutions, associated with different equilibrium points in state space: 
Class A solutions are connected with the M points and are described by two 
parameters. Class B solutions are connected with the $\tilde{\rm M}$
points and are described by one parameter. Both of these subclasses
contain two different families of solutions: 
(1) singular solutions (K--M, K--$\tilde{\rm M}$ orbits); and 
(2) regular solutions (C--M, C--$\tilde{\rm M}$ orbits).
Only the latter contain a sonic point. 
All these types of solutions are illustrated in Figs. \ref{fig:mgnu} and 
\ref{fig:mcc}, where the arrows indicate whether solutions are
future asymptotic to M or $\tilde{\rm M}$.

Solutions in class A have $V\rightarrow1$, $S\rightarrow\infty$, 
$\mu t^2\rightarrow0$ and $2m/R\rightarrow0$ at some {\em finite}
value $z=z_*$. Although this limit is reached at finite $z$, 
it should be pointed out that most investigations of the critical
solution use the physical distance $R$, which is infinite.
Examples of such solutions are illustrated by the dotted curves 
in Figs. \ref{fig:mgnu} and \ref{fig:mcc}. Solutions in class B have 
$V\rightarrow V_*>1$, $S\rightarrow\infty$, $\mu t^2\rightarrow0$ and 
$2m/R\rightarrow0$ as $z\rightarrow\infty$. Examples of these solutions
are represented by the dashed curves in Figs. \ref{fig:mgnu} and
\ref{fig:mcc}. Both classes are asymptotically dispersive and
solutions asymptotic to K points also have $S\rightarrow0$ at a finite
value of $z$, indicating the formation of a singularity in the past
(assuming the time direction indicated in the figures). Even though
$2m/R\rightarrow0$ for both classes, the mass $m$ need not vanish. 

It should be emphasized that these solutions are not asymptotically flat
in the usual global sense, in which there is a certain radial decay of the 
curvature towards spatial infinity, see, e.g.,  \cite{Bartnik1986}. We 
are not considering an isolated system here but rather a fluid spacetime
in which the Minkowski geometry is obtained asymptotically along certain 
coordinate lines. It can be shown that the curvature vanishes asymptotically 
as the M ($\tilde{\rm M}$) point is approached and $r\rightarrow\infty$ 
(and hence $t\rightarrow\infty$). Because the fluid becomes infinitely 
diluted, the situation is analogous to that of the open Friedmann
solution, in which the Milne solution is approached asymptotically
along certain time-lines at late times
(see, e.g., \cite{book:WainwrightEllis1997}). 

\section{The critical solution}\label{sec:crit}

Critical phenomena in gravitational collapse were first studied by
Choptuik \cite{Choptuik1993} and remain an active field of research, see,
e.g., \cite{Gundlach1998} and references therein. The solution at
the threshold of black-hole formation in spherically symmetric
radiation fluid collapse, corresponding to $\alpha=\case13$, was
studied by Evans \& Coleman \cite{EvansColeman1994}. They found it
to be a self-similar solution distinguished by the following criteria:

(1) It is everywhere analytic, or at least $C^\infty$. In
particular it has a regular center, and also crosses the sonic surface
in an analytic way.

(2) It has a collapsing interior surrounded by an expanding 
exterior. This means that the radial fluid three-velocity $V_R$ 
associated with a Schwarzschild foliation (which is different from
the function $V$) has exactly one zero.

Subsequently, other authors \cite{Maison1996,Koike-et-al1999}
have used these criteria to investigate the critical solution for 
other values of $\alpha$. For a recent review, see \cite{Gundlach1998}. 

The uniqueness of the critical solution can be understood as follows:
For each value of the equation-of-state parameter $\alpha$, there
exists a one-parameter set of solutions with a regular center and a
one-parameter set of solutions analytic at the sonic line
(\cite{CarrColey1998SS}). Thus it is not surprising that the first
condition leads to a discrete set of solutions. There is only one
solution in this set that satisfies the second condition and this is
the critical solution. 

\begin{figure}
  \centerline{\hbox{\epsfig{figure=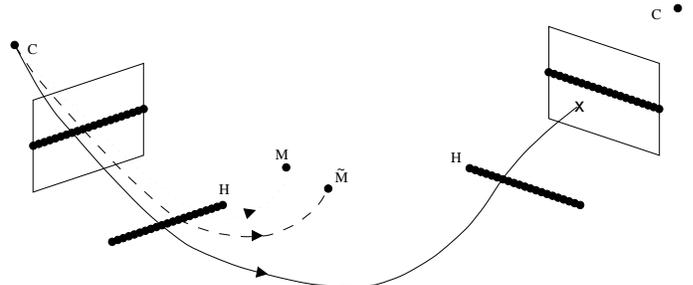,width=0.5\textwidth}}}
  \caption{Critical orbits in the state space. Arrows go from the regular 
    center towards the infinitely dispersed state. Note that the
    orbits correspond to different equations of state, i.e., different
    values of $\alpha$, and thus belong to different state
    spaces. The three orbits exemplify the cases
    $0<\alpha\lesssim0.28$ (full curve), $\alpha\approx0.28$
    (dashed), and $0.28\lesssim\alpha<1$ (dotted).}\label{fig:critgnu}
\end{figure}

\begin{figure}
  \centerline{\hbox{\epsfig{figure=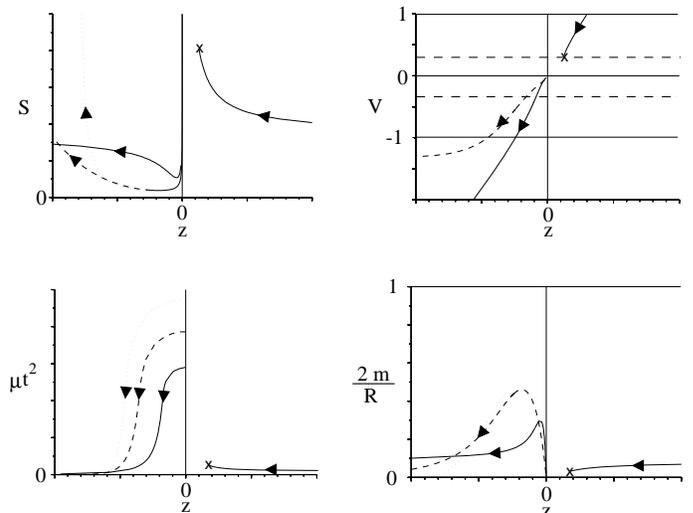,width=0.5\textwidth}}}
  \caption{Physical quantities for the critical solution. 
    The dash-dotted lines in the $V(z)$ diagram correspond to the sonic
    surfaces. Other designations are given in the caption of 
    Fig. \ref{fig:critgnu}. Note that for $\alpha\gtrsim0.28$ (dotted),
    the critical solution tends to a finite value of $z$.}\label{fig:critcc}
\end{figure}

We now examine the critical solution in terms of both the state space
of the self-similar spherically symmetric perfect-fluid solutions and
the behavior of the various physical quantities. The results are
summarized in Figs. \ref{fig:critgnu} and \ref{fig:critcc}.

Starting from the regular center C, a numerical investigation shows that
for all equations of state, the orbit of the critical solution passes
through the sonic line and enters the spatially self-similar region
(with $|V|>1$). It turns out that for $\alpha$ in the range
$0<\alpha\lesssim0.28$, it is of the asymptotically quasi-static kind:
it passes through the spatially self-similar region and enters a
second timelike self-similar region, finally reaching another sonic
point (indicated by `{\sf x}' in the figures), which is generally
irregular. However, this does not invalidate the solution as being the
critical one, since the solution describing the inner collapsing
region is usually matched to an asymptotically flat exterior geometry
sufficiently far from the center. An example of a solution belonging
to this class is given by the full curves in
Figs. \ref{fig:critgnu} and \ref{fig:critcc}. 

We find that for the limiting case $\alpha\approx0.28$, the critical
solution is an asymptotically Minkowski solution of class B, whose
orbit ends at an equilibrium point $\tilde{\rm M}$. In
Figs. \ref{fig:critgnu} and \ref{fig:critcc}, this solution
corresponds to the dashed lines. For $0.28\lesssim\alpha<1$, we find
that the critical solution belongs to the asymptotically Minkowski
solutions of class A, whose orbit ends at an equilibrium point M.
A solution representing this class is indicated by the dotted curves in
Figs. \ref{fig:critgnu} and \ref{fig:critcc}.

For $\alpha\geq\alpha_*\approx0.89$, the investigations in 
\cite{Goliath-et-al1998TSS,Maison1996,Koike-et-al1999}
indicate that the critical solution is already irregular at the
first sonic point. As the matching must be performed outside the sonic
point \cite{OriPiran1990,FoglizzoHenriksen1993}, the solution would 
then be unphysical.  However, Neilsen \& Choptuik
\cite{NeilsenChoptuik1998} have recently demonstrated the
existence of a regular critical solution for $\alpha\geq\alpha_*$ as well.
Our present investigation supports their analysis. 

To understand what happens for $\alpha=\alpha_*$, we consider
equations of state near this value. It turns out that the behavior
can be understood in terms of the stability near the sonic line. In
order for a solution to be regular at the sonic surface, the
corresponding orbit must approach the sonic line along one of (at
most) two possible directions. Each of these directions is 
associated with an eigenvalue -- the direction corresponding to the
smaller eigenvalue is called {\em dominant} and is associated with a
one-parameter family of solutions (containing just one $C^\infty$
solution), the other is called {\em secondary} and is associated with
an isolated solution. For $\alpha<\alpha_*$, the critical solution
corresponds to the secondary direction. However, when $\alpha$ gets
close to $\alpha_*$, the eigenvalue associated with the critical
solution approaches that of the other direction. 
For $\alpha=\alpha_*$, the eigenvalues (and directions) are equal, 
corresponding to a {\em degenerate node}, and for $\alpha>\alpha_*$, the 
critical solution is associated with the dominant direction. Thus a 
transition from the secondary direction to the dominant direction occurs
at $\alpha=\alpha_*$. This transition results in severe numerical
difficulties, so very high numerical precision is required to
investigate such solutions, as pointed out in \cite{NeilsenChoptuik1998}. 

\section{Conclusions}

Our key predictions can be summarized as follows:

(1) When one gets sufficiently close to the critical solution that the
large-$z$ behavior can be studied, then this solution should have the
various asymptotic features we predict for different values of $\alpha$. 

(2) There should be a sudden transition in the nature of the critical
solution as $\alpha$ passes through 0.28, with the solution going from
the asymptotically quasi-static form to asymptotically Minkowski
form. However, it should be emphasized that the solution is only flat
towards null infinity for $\alpha>1/3$, so one still needs to match to a
non-self-similar region on a spacelike surface. 

(3) Although the asymptotically flat limit is reached at finite $z$, most
critical workers use the physical distance, which is infinite. However, it 
should be pointed out that in the stiff case ($\alpha=1$), the
asymptotically flat state is reached at {\it finite} physical
distance, which should lead to some anomalies, see
\cite{CarrColey1999classi}).

(4) Although we have not explained why the critical solution is analytic
at the sonic point (this presumably relates to the usual stability
criterion), we have used the asymptotic features to explain why the
analytic solution is unique for given $\alpha$. The relationship to
solutions which are regular but not analytic at the sonic point is
discussed in more detail by Carr \& Henriksen \cite{CarrHenriksen1999}.


\begin{references}

  \bibitem{Choptuik1993}
  M. W. Choptuik,
  Phys. Rev. Lett. {\bf 70}, 9 (1993).

  \bibitem{AbrahamsEvans1993}
  A. M. Abrahams and C. R. Evans,
  Phys. Rev. Lett. {\bf 70}, 2980 (1993).

  \bibitem{Choptuik1994}
  M. W. Choptuik,
  in \emph{Deterministic chaos in general relativity}, eds. D. Hobill
  et al., Plenum Press, New York (1994).

  \bibitem{CarrColey1998SS}
  B. J. Carr and A. A. Coley,
  Class. Quant. Grav. {\bf 16}, R1 (1999).

  \bibitem{EvansColeman1994}
  C. R. Evans and J. S. Coleman,
  Phys. Rev. Lett. {\bf 72}, 1782 (1994).

  \bibitem{Gundlach1995}
  C. Gundlach,
  Phys. Rev. Lett. {\bf 75}, 3214 (1995).

  \bibitem{Maison1996}
  D. Maison,
  Phys. Lett. B {\bf 366}, 82 (1996).

  \bibitem{Koike-et-al1999}
  T. Koike, T. Hara, and S. Adachi,
  Phys. Rev. D {\bf 59}, 104008 (1999).

  \bibitem{Frolov1997}
  V. Frolov,
  Phys. Rev. D {\bf 56} 6433 (1997).
 
  \bibitem{CarrColey1999classi}
  B. J. Carr and A. A. Coley,
  ``A complete classification of spherically symmetric perfect fluid
  similarity solutions'', 
  preprint gr-qc/9901050 (1999).

  \bibitem{Goliath-et-al1998SSS}
  M. Goliath, U. S. Nilsson, and C. Uggla,
  Class. Quant. Grav. {\bf 15}, 167 (1998).

  \bibitem{Goliath-et-al1998TSS}
  M. Goliath, U. S. Nilsson, and C. Uggla,
  Class. Quant. Grav. {\bf 15}, 2841 (1998).

  \bibitem{book:WainwrightEllis1997}
  J. Wainwright and G. F. R. Ellis, {\em dynamical systems in cosmology}
  (Cambridge University Press, Cambridge, 1997).

  \bibitem{Carr-et-al1999}
  B. J. Carr, A. A. Coley, M. Goliath, U. S. Nilsson, and C. Uggla,
  preprint gr-qc/9902070 (1999).

  \bibitem{CahillTaub}
  M. E. Cahill and A. H. Taub,
  Commun. Math. Phys. {\bf 21}, 1 (1971).
  
  \bibitem{OriPiran1990}
  A. Ori and T. Piran,
  Phys. Rev. D {\bf 42}, 1068 (1990).

  \bibitem{HernandezMisner1966}
  W. C. Hernandez and C. W. Misner,
  Astrophys. J. {\bf 143}, 452 (1966).

  \bibitem{Tolman1934}
  R. C. Tolman,
  Proc. Nat. Acad. Sci. {\bf 20}, 169 (1934).
  
  \bibitem{FoglizzoHenriksen1993}
  T. Foglizzo and R. N. Henriksen,
  Phys. Rev. D {\bf 48}, 4645 (1993).

  \bibitem{Bartnik1986}
  R. Bartnik, Commun. Pure App. Math. {\bf 39}, 661 (1986).
  
  \bibitem{Gundlach1998}
  C. Gundlach,
  Adv. Theor. Math. Phys. {\bf 2}, 1 (1998).
  
  \bibitem{NeilsenChoptuik1998}
  D. W. Neilsen and M. W. Choptuik,
  ``Critical phenomena in perfect fluids'',
  preprint gr-qc/9812053 (1998).

  \bibitem{CarrHenriksen1999}
  B. J. Carr and R. N. Henriksen,
  in preparation.
  
\end{references}
\end{document}